\documentclass[twocolumn,showpacs,aps,prd,superscriptaddress,floatfix]{revtex4}

\usepackage{graphicx}
\usepackage{verbatim}
\usepackage{dcolumn}
\usepackage{amsmath}
\usepackage{epsfig}
\usepackage{subfigure}

\newcommand{\BaBarYear}       {10}
\newcommand{\BaBarNumber}     {}
\newcommand{\SLACPubNumber} {}

\newcommand{\BaBarType}      {PUB}  % Journal publication

\RequirePackage{xspace}

\usepackage{relsize}
\def\babar{\mbox{\slshape B\kern-0.1em{\smaller A}\kern-0.1em
    B\kern-0.1em{\smaller A\kern-0.2em R}}}

\def\en         {\ensuremath{e^-}\xspace}   % electron negative (\em is taken)
\def\ep         {\ensuremath{e^+}\xspace}
 
\def\epem       {\ensuremath{e^+e^-}\xspace}

\def\mup        {\ensuremath{\mu^+}\xspace}
\def\mun        {\ensuremath{\mu^-}\xspace} % muon negative (\mum is taken)
\def\mumu       {\ensuremath{\mu^+\mu^-}\xspace}

\def\taup       {\ensuremath{\tau^+}\xspace}
\def\taum       {\ensuremath{\tau^-}\xspace}
\def\tautau     {\ensuremath{\tau^+\tau^-}\xspace}

\def\ellell     {\ensuremath{\ell^+ \ell^-}\xspace}

\def\nub        {\ensuremath{\overline{\nu}}\xspace}
\def\nunub      {\ensuremath{\nu{\overline{\nu}}}\xspace}
\def\nub        {\ensuremath{\overline{\nu}}\xspace}
\def\nunub      {\ensuremath{\nu{\overline{\nu}}}\xspace}
\def\nue        {\ensuremath{\nu_e}\xspace}
\def\nueb       {\ensuremath{\nub_e}\xspace}

\def\num        {\ensuremath{\nu_\mu}\xspace}
\def\numb       {\ensuremath{\nub_\mu}\xspace}

\def\nut        {\ensuremath{\nu_\tau}\xspace}
\def\nutb       {\ensuremath{\nub_\tau}\xspace}

\def\nul        {\ensuremath{\nu_\ell}\xspace}
\def\nulb       {\ensuremath{\nub_\ell}\xspace}

\def\g     {\ensuremath{\gamma}\xspace}
\def\gaga  {\ensuremath{\gamma\gamma}\xspace}  %% changed from \gg, which is >>
\def\ggstar{\ensuremath{\gamma\gamma^*}\xspace}

\def\piz   {\ensuremath{\pi^0}\xspace}

\def\ppz   {\ensuremath{\pi^0\pi^0}\xspace}
\def\pip   {\ensuremath{\pi^+}\xspace}
\def\pim   {\ensuremath{\pi^-}\xspace}
\def\pipi  {\ensuremath{\pi^+\pi^-}\xspace}

\def\pimp  {\ensuremath{\pi^\mp}\xspace}

\def\kaon  {\ensuremath{K}\xspace}
\def\Kbar  {\kern 0.2em\overline{\kern -0.2em K}{}\xspace}

\def\Kz    {\ensuremath{K^0}\xspace}
\def\Kzb   {\ensuremath{\Kbar^0}\xspace}
\def\KzKzb {\ensuremath{\Kz \kern -0.16em \Kzb}\xspace}
\def\Kp    {\ensuremath{K^+}\xspace}
\def\Km    {\ensuremath{K^-}\xspace}
\def\Kpm   {\ensuremath{K^\pm}\xspace}

\def\KpKm  {\ensuremath{\Kp \kern -0.16em \Km}\xspace}
\def\KS    {\ensuremath{K^0_{\scriptscriptstyle S}}\xspace} 
\def\KL    {\ensuremath{K^0_{\scriptscriptstyle L}}\xspace} 
\def\Kstarz  {\ensuremath{K^{*0}}\xspace}

\def\Kstar   {\ensuremath{K^*}\xspace}

\def\Kstarp  {\ensuremath{K^{*+}}\xspace}

\newcommand{\etapr}{\ensuremath{\eta^{\prime}}\xspace}

\def\Dbar    {\kern 0.2em\overline{\kern -0.2em D}{}\xspace}

\def\Dz      {\ensuremath{D^0}\xspace}
\def\Dzb     {\ensuremath{\Dbar^0}\xspace}
\def\DzDzb   {\ensuremath{\Dz {\kern -0.16em \Dzb}}\xspace}
\def\Dp      {\ensuremath{D^+}\xspace}
\def\Dm      {\ensuremath{D^-}\xspace}

\def\DpDm    {\ensuremath{\Dp {\kern -0.16em \Dm}}\xspace}
\def\Dstar   {\ensuremath{D^*}\xspace}

\def\Dstarz  {\ensuremath{D^{*0}}\xspace}

\def\Dstarp  {\ensuremath{D^{*+}}\xspace}
\def\Dstarm  {\ensuremath{D^{*-}}\xspace}

\newcommand{\dstr}{\ensuremath{\Dstar}\xspace}

\def\B       {\ensuremath{B}\xspace}
\def\Bbar    {\kern 0.18em\overline{\kern -0.18em B}{}\xspace}

\def\BB      {\ensuremath{B\Bbar}\xspace} 
\def\Bz      {\ensuremath{B^0}\xspace}
\def\Bzb     {\ensuremath{\Bbar^0}\xspace}
\def\BzBzb   {\ensuremath{\Bz {\kern -0.16em \Bzb}}\xspace}
\def\Bu      {\ensuremath{B^+}\xspace}
\def\Bub     {\ensuremath{B^-}\xspace}
\def\Bp      {\ensuremath{\Bu}\xspace}

\def\Bpm     {\ensuremath{B^\pm}\xspace}

\def\BpBm    {\ensuremath{\Bu {\kern -0.16em \Bub}}\xspace}

\def\BorBbar    {\kern 0.18em\optbar{\kern -0.18em B}{}\xspace}
\def\DorDbar    {\kern 0.18em\optbar{\kern -0.18em D}{}\xspace}
\def\KorKbar    {\kern 0.18em\optbar{\kern -0.18em K}{}\xspace}

\def\jpsi     {\ensuremath{{J\mskip -3mu/\mskip -2mu\psi\mskip 2mu}}\xspace}

\mathchardef\Upsilon="7107
\def\Y#1S{\ensuremath{\Upsilon{(#1S)}}\xspace}% no space before {...}!

\def\FourS {\Y4S}

\mathchardef\Deltares="7101
\mathchardef\Xi="7104
\mathchardef\Lambda="7103
\mathchardef\Sigma="7106
\mathchardef\Omega="710A

\def\Deltabar{\kern 0.25em\overline{\kern -0.25em \Deltares}{}\xspace}
\def\Lbar{\kern 0.2em\overline{\kern -0.2em\Lambda\kern 0.05em}\kern-0.05em{}\xspace}
\def\Sigbar{\kern 0.2em\overline{\kern -0.2em \Sigma}{}\xspace}
\def\Xibar{\kern 0.2em\overline{\kern -0.2em \Xi}{}\xspace}
\def\Obar{\kern 0.2em\overline{\kern -0.2em \Omega}{}\xspace}
\def\Nbar{\kern 0.2em\overline{\kern -0.2em N}{}\xspace}
\def\Xb{\kern 0.2em\overline{\kern -0.2em X}{}\xspace}

\def\X {\ensuremath{X}\xspace}

\def\BR         {{\ensuremath{\cal B}\xspace}}

\def\bpsikl     {\ensuremath{\Bz \to \jpsi \KL}\xspace}

\def\Bzbtox     {\ensuremath{\Bzb \to \X}\xspace}

\def\upsbzbz {\ensuremath{\FourS \to \BzBzb}\xspace}
\def\upsbpbm {\ensuremath{\FourS \to \BpBm}\xspace}

\newcommand{\tev}{\ensuremath{\mathrm{\,Te\kern -0.1em V}}\xspace}
\newcommand{\gev}{\ensuremath{\mathrm{\,Ge\kern -0.1em V}}\xspace}
\newcommand{\mev}{\ensuremath{\mathrm{\,Me\kern -0.1em V}}\xspace}
\newcommand{\kev}{\ensuremath{\mathrm{\,ke\kern -0.1em V}}\xspace}
\newcommand{\ev}{\ensuremath{\mathrm{\,e\kern -0.1em V}}\xspace}
\newcommand{\gevc}{\ensuremath{{\mathrm{\,Ge\kern -0.1em V\!/}c}}\xspace}
\newcommand{\mevc}{\ensuremath{{\mathrm{\,Me\kern -0.1em V\!/}c}}\xspace}
\newcommand{\gevcc}{\ensuremath{{\mathrm{\,Ge\kern -0.1em V\!/}c^2}}\xspace}
\newcommand{\mevcc}{\ensuremath{{\mathrm{\,Me\kern -0.1em V\!/}c^2}}\xspace}
\def\invfb   {\ensuremath{\mbox{\,fb}^{-1}}\xspace}

\def\mus  {\ensuremath{\rm \,\mus}\xspace}

\def\mus        {\ensuremath{\,\mu{\rm s}}\xspace}    %% microsecond
\def\degrees{\ensuremath{^{\circ}}\xspace}
\def\to                 {\ensuremath{\rightarrow}\xspace}

\def\pep2{PEP-II}
\def\BF{$B$ Factory}

\def\gsim{{~\raise.15em\hbox{$>$}\kern-.85em
          \lower.35em\hbox{$\sim$}~}\xspace}
\def\lsim{{~\raise.15em\hbox{$<$}\kern-.85em
          \lower.35em\hbox{$\sim$}~}\xspace}

\def\epstag  {\ensuremath{\varepsilon_{\rm tag}}\xspace}

\xspace

\newcommand{\epjBase}        {Eur.\ Phys.\ Jour.\xspace}
\newcommand{\jprlBase}       {Phys.\ Rev.\ Lett.\xspace}
\newcommand{\jprBase}        {Phys.\ Rev.\xspace}
\newcommand{\jplBase}        {Phys.\ Lett.\xspace}
\newcommand{\nimBaseA}       {Nucl.\ Instrum.\ Methods Phys.\ Res., Sect.\ A\xspace}
\newcommand{\nimBaseB}       {Nucl.\ Instrum.\ Methods Phys.\ Res., Sect.\ B\xspace}
\newcommand{\nimBaseC}       {Nucl.\ Instrum.\ Methods Phys.\ Res., Sect.\ C\xspace}
\newcommand{\nimBaseD}       {Nucl.\ Instrum.\ Methods Phys.\ Res., Sect.\ D\xspace}
\newcommand{\nimBaseE}       {Nucl.\ Instrum.\ Methods Phys.\ Res., Sect.\ E\xspace}
\newcommand{\npBase}         {Nucl.\ Phys.\xspace}
\newcommand{\zpBase}         {Z.\ Phys.\xspace}

\def\jetset74   {\mbox{\tt Jetset \hspace{-0.5em}7.\hspace{-0.2em}4}\xspace}
\newcommand{\gevcccc}{\ensuremath{{\mathrm{\,Ge\kern -0.1em V^2\!/}c^4}}\xspace}

\def\K {\ensuremath{K}\xspace}

\def\modekJpsill {\ensuremath{\Bp \to K^+ J/\psi(\to \ellell)}\xspace}

\def\modekavgnn {\ensuremath{B \to K\nunub}\xspace}

\def\modekplnn {\ensuremath{B^{+}\to K^{+}\nunub}\xspace}

\def\BBbar {\ensuremath{B\Bbar}}
\def\modekzeroallnn {\ensuremath{B^{0}\to K^{0}\nunub}\xspace}
\def\Bsig {\ensuremath{B_{\rm roe}}}
\def\Btag {\ensuremath{B_{\rm rec}}}
\def\Eex {\ensuremath{E_{\rm extra}}}
\def\mnn {\ensuremath{q^{2}}}

\def\figurebox#1#2#3{%
    \def\arg{#3}%
    \ifx\arg\empty
    {\hfill\vbox{\hsize#2\hrule\hbox to #2{\vrule\hfill\vbox to #1{\hsize#2\vfill}\vrule}\hrule}\hfill}%
    \else
    {\hfill\epsfbox{#3}\hfill}%
    \fi}

\usepackage[normalem]{ulem}

\begin{document}

\begin{flushleft}
\babar-\BaBarType-\BaBarYear/011 \BaBarNumber \\
SLAC-PUB-14237\SLACPubNumber\\
\end{flushleft}

 % Title of the paper
\title{

  {\large \bf \boldmath
Search for the Rare Decay \modekavgnn
}
}

%BABAR author list
%\input{authors_may2010.tex}
%% author list as of 03-May-2010 (442 authors)
%
\author{P.~del~Amo~Sanchez}
\author{J.~P.~Lees}
\author{V.~Poireau}
\author{E.~Prencipe}
\author{V.~Tisserand}
\affiliation{Laboratoire d'Annecy-le-Vieux de Physique des Particules (LAPP), Universit\'e de Savoie, CNRS/IN2P3,  F-74941 Annecy-Le-Vieux, France}
\author{J.~Garra~Tico}
\author{E.~Grauges}
\affiliation{Universitat de Barcelona, Facultat de Fisica, Departament ECM, E-08028 Barcelona, Spain }
\author{M.~Martinelli$^{ab}$}
\author{A.~Palano$^{ab}$ }
\author{M.~Pappagallo$^{ab}$ }
\affiliation{INFN Sezione di Bari$^{a}$; Dipartimento di Fisica, Universit\`a di Bari$^{b}$, I-70126 Bari, Italy }
% ...
\collaboration{The \babar\ Collaboration}
\noaffiliation

\date{\today}

% Abstract
\begin{abstract}
\noindent
We present a search for the rare decays $\modekplnn$ and $\modekzeroallnn$
using $459$ million $\BB$ pairs collected with the \babar\ detector
at the SLAC    National Accelerator Laboratory.
Flavor-changing neutral-current decays such as these are forbidden at tree level
but can occur through one-loop diagrams in the Standard Model (SM), with possible contributions from new physics  at the same
order.
The presence of two neutrinos in the final state makes identification of
signal events challenging, so reconstruction  in
the semileptonic decay channels $\B \to D^{(*)} l \nu$ of the  $B$ meson recoiling from the signal $B$  is used to suppress backgrounds.
We set an upper limit at the 90\% confidence level of $1.3 \times 10^{-5}$ on the total branching fraction for \modekplnn,
and $5.6 \times 10^{-5}$ for $\modekzeroallnn$. We additionally report 90\% confidence level upper limits on partial branching fractions in two ranges
of di-neutrino mass squared for \modekplnn.

\end{abstract}

% PACS, the Physics and Astronomy
                             % Classification Scheme.

\maketitle

\begin{figure}[b!]
\centering
\includegraphics[scale=0.35, clip, trim= 2.3in 3.2in 0.8in 3.2in]{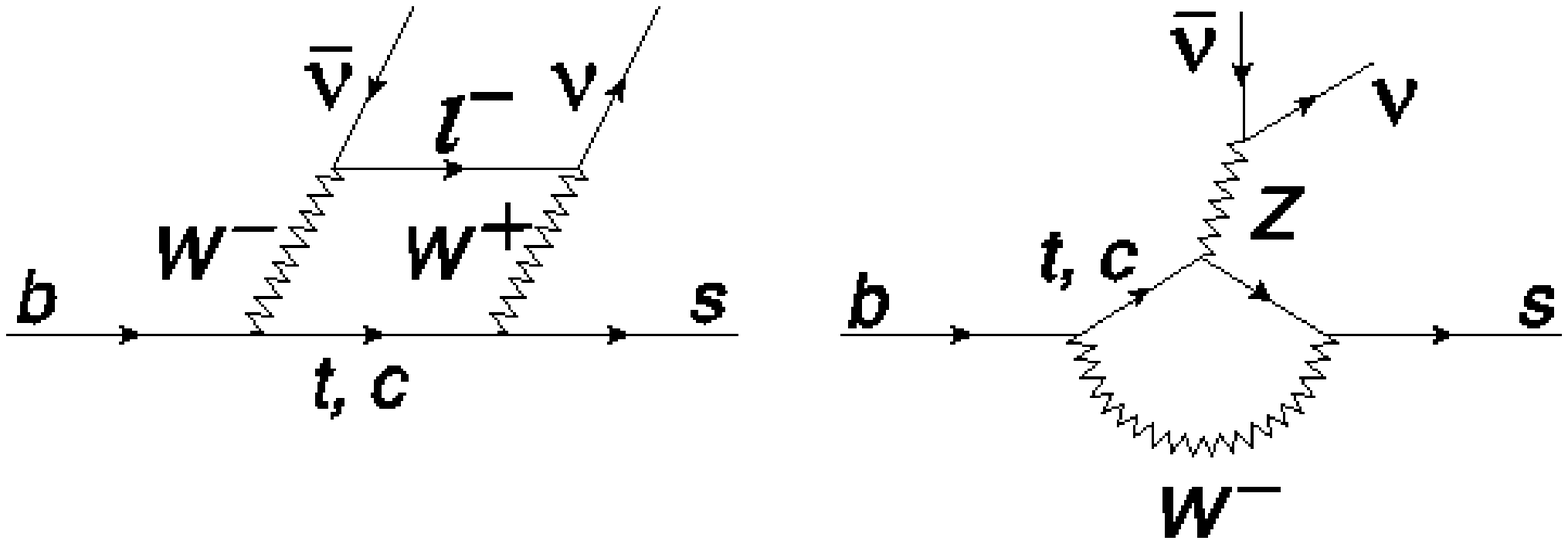}
\\
\caption{Lowest-order Feynman diagrams for $\modekavgnn$, with the $W$ box on the left and $Z$ penguin on the right.}
\label{fig:sll_diagrams}
\end{figure}

% Paper text
\clearpage
The decays $\modekavgnn$ arise from flavor-changing neutral
currents (FCNC), which are forbidden
at tree level in the Standard Model (SM).
The lowest-order SM processes contributing to these decays
are the $W$ box and the $Z$ penguin diagrams
shown in Fig.~\ref{fig:sll_diagrams}. New physics contributions
may enter  at the same order as the
SM. These contributions, some of which could increase the branching fraction by up to ten times
relative to the SM, include:
unparticle models~\cite{Aliev:2007gr},
Minimal Supersymmetric extension of the Standard Model at large $\tan \beta$ ~\cite{Yamada:2007me},
models with a single universal extra dimension~\cite{Colangelo:2006vm},
scalar Weakly Interacting Massive Particle (WIMP) dark matter~\cite{Bird:2004ts} and WIMP-less dark matter~\cite{McKeen:2009rm}.
A recent SM prediction (ABSW model~\cite{Altmannshofer:2009ma}) for the total $\modekavgnn$ branching fraction is
$(4.5 \pm 0.7) \times 10^{-6}$, while an earlier
 prediction  (BHI model~\cite{Buchalla:2000sk}), based on a different form factor model, is $(3.8^{+1.2}_{-0.6} )\times 10^{-6}$.
The BHI model was used by previous analyses~\cite{belle:2007zk,Aubert:2004ws} and provides a baseline for comparison between results. The current experimental upper limit (UL) on the total branching fraction for $\modekplnn$~(charge conjugation is implied throughout) is
$1.4\times 10^{-5}$ at the 90\% confidence level (CL) from the Belle Collaboration~\cite{belle:2007zk},
while an earlier \babar\ analysis set an
UL  of $5.2\times 10^{-5}$ (90\% CL)~\cite{Aubert:2004ws}.
 The only existing UL on the total branching fraction for $\modekzeroallnn$ is
$1.6\times 10^{-4}$ (90\% CL) from Belle~\cite{belle:2007zk}.

We report results of a search for $\modekplnn$ and $\modekzeroallnn$, with branching fractions for both decays as well as for the combination $\modekavgnn$. We also report on partial
branching fractions for $\modekplnn$ in two regions of di-neutrino invariant mass squared ($\mnn$). The low-$\mnn$ region ($\mnn < 0.4 m^{2}_{B}$)
is selected by requiring $p^{*}_{\Kp}>1.5\gevc$  and the high-$\mnn$ region ($\mnn > 0.4 m^{2}_{B}$)
by $p^{*}_{\Kp}<1.5\gevc$ in the $\FourS$ center-of-mass system (CMS)~\cite{asterisk}, where $m_{B}$ is the mass of the $B$ meson and $p^{*}_{\Kp}$ is the magnitude of the CMS 3-momentum of the signal $\Kp$ candidate. The high-$\mnn$ region is of theoretical interest because the partial branching fraction in this region could be enhanced under some new physics
models~\cite{Altmannshofer:2009ma}.

This analysis is based on a data sample of $(459.0 \pm 5.1) \times 10^{6}$ $\BBbar$ pairs, corresponding to an integrated luminosity of $\sim 418 \invfb$ of $\epem$ colliding-beam data and recorded at
the $\FourS$ resonance with the \babar\ detector~\cite{Aubert:2001tu} at the SLAC \hbox{PEP-II} asymmetric-energy $B$ Factory.  Charged particle tracking is provided by a five-layer silicon vertex tracker
and a 40-layer drift chamber in a 1.5 T magnetic field.
A CsI(Tl) electromagnetic calorimeter (EMC) is used to measure photon
energies and directions and to identify
electrons.  All quantities in this paper which are measured by the EMC are required to exceed a minimum
20 MeV cluster energy, unless a higher threshold is explicitly noted. 
The magnetic flux return from the solenoid, instrumented with resistive plate chambers and limited streamer tubes (IFR),  provides  muon identification. We identify $\Kp$ candidates by using
 a detector of internally reflected Cherenkov light (DIRC) as well as ionization energy loss information from the
tracking system.

Due to the presence of two neutrinos in the $\modekavgnn$ final state,
it is not possible to exploit the kinematic constraints on the $B$ mass and energy which are
typically used to distinguish signal and background events
in $B$ meson decays at the $\FourS$. Instead, before looking for the signal decay, we  first reconstruct
a  $B$ decay ($\Btag$) in one of several exclusive $D^{(*)}l\nu$ semileptonic final states. We then search for the signal $\modekavgnn$ among the remaining charged and neutral particles in the detector that are not part of 
the $\Btag$.  We collectively refer to these remaining particles as  $\Bsig$ for  ``rest of the event."
This strategy is common to several \babar\ analyses~\cite{Aubert:2007bx,Aubert:2009rk} and  has the advantage of higher efficiency compared with reconstruction of the $\Btag$ in hadronic decay modes~\cite{Aubert:2004ws}. 

We reconstruct the $D$ candidates in the following decay modes:  $\Km \pip$, $\Km \pip\pip$, $\Km \pip\pip\pim$, $\Km \pip \piz$, $\KS\pip$, and $\KS\pip\pim$.  The $\KS$ candidates, reconstructed in the  $\KS \to \pi^{+}\pi^{-}$ mode, are required to have a
  $\pipi$ invariant mass within $25\mevcc$ of the nominal $\KS$ mass.  $D$ candidates are similarly required to have a reconstructed invariant mass
within $60 \mevcc$ of the nominal value~\cite{pdg:2008}, except for the $\Km \pi^{+}\pi^{0}$ mode where the range is $100 \mevcc$.  We  form $D^{*0}\to D^{0} \pi^{0}$, $D^{*+}\to D^{0} \pi^{+}$, and $D^{*+}\to D^{+}\pi^{0}$candidates with a required mass difference $(m(D^{*})-m(D))$ in the range 130-170 $\mevcc$.  In addition, we combine $D$ and $\gamma$ candidates to form  $D^{*}$ candidates with a required mass difference in the range 120-170  $\mevcc$.  A $D^{(*)}$ candidate is combined with an identified electron or muon with momentum above 0.8 GeV/c in the CMS  to form a $\Btag$ candidate. In events with multiple
reconstructed $\Btag$ candidates, we select the candidate with the highest probability that the daughter tracks originate from a common vertex.
After a $\Btag$ candidate has been identified, the remaining charged and
neutral decay products are used to classify the $\Bsig$ as either
a background event or a possible signal candidate.

As a first step in refining the selection of $\Bsig$ candidates, we veto $K$ candidates which, when combined with a remaining charged or neutral pion
candidate, have
a $K\pi$ invariant mass within $75\mevcc$ of the nominal $K^{*}(892)$ mass.
We also veto events where  a  remaining charged track can be
combined with a $\piz$ candidate to yield a $\rho^{+}$ candidate, with a mass window $0.45 < m(\rho^{+}) < 1.10 \gevcc$. Similarly vetoed are events where three remaining charged tracks can be combined to yield an $a_{1}^{+}$ candidate, with a mass window $0.6 < m(a_{1}^{+}) < 2.0 \gevcc$.
These vetoes eliminate, with little loss of signal efficiency, sizable backgrounds that consist mostly of random track combinations. 
 After the vetoes, $\Bp$ ($\Bz$) signal candidate events are required to possess $\Kp$ ($\KS\to\pipi$) candidates,
accompanied by at most two (one) additional charged tracks, which are assumed to have been incorrectly left out of the $\Btag$.  
For the $\Kp$ final state, the
$\Btag$ lepton daughter and the $\Kp$ are also required to be oppositely charged. For the $\KS$ final state, signal candidates are required to have a  $\pipi$ invariant mass within $25\mevcc$ of the nominal $\KS$ mass.

At this stage of the selection,  each event has a $\Btag$ candidate representing a $B$ meson reconstructed in a semileptonic decay and a $\Bsig$ candidate formed from the rest of the event, with the latter representing the signal decay. In simulated $\Kp$ ($\KS$) signal events that have passed this selection,  99\% (92\%) of events have a correctly identified signal  $\Kp$ ($\KS$). However, a large background still remains.  Further background suppression is achieved using 
 a multivariate event selection algorithm, a bagged decision tree (BDT)~\cite{Breiman,Narsky:2005hn}, that can 
leverage many weak discriminating variables to achieve high background rejection.  Such an algorithm needs to be trained with
simulated signal and background events, henceforth referred to as  Monte Carlo (MC) events.
We use a GEANT4~\cite{Agostinelli:2002hh} detector simulation to obtain large samples
of simulated  signal events generated with a pure phase-space model (which are later re-scaled to the BHI signal model),
as well as  samples of non-resonant $e^{+}e^{-}\to q \bar{q}$ $(q=u, d, s, c)$, $\BBbar$, and $\tautau$ background
events, whose sizes are one ($uds$), two ($c \bar{c}$), three ($\BBbar$), and one ($\tautau$) times luminosity. These background events
are augmented with a separate sample, with a size 13 times luminosity, of simulated $\BBbar$ doubly semileptonic events,
the largest source of background.

We construct  two ensembles of BDTs, one for the $\Kp$ signal mode and one for the  $\KS$. To create an ensemble, 
we repeatedly  divide the total signal and  background
datasets in half randomly, creating 20 distinct BDT training and validation datasets,
where each dataset has a $50\%$ correlation with any other because approximately 50\% of the events are shared.
This procedure makes optimal use of the limited statistics of MC events that pass the initial event
selection and results in a more statistically precise unbiased estimate of background contributions.
Use of the ensemble of 20~BDTs created for each final state also averages out the variations in BDT response compared
to a single BDT trained and validated with a single division of the simulated
signal and background datasets \cite{Bootstrap, Barlow}.  The choice of 20 divisions, instead of a lower 
or higher number, represents a balance between minimizing the variation versus minimizing the overhead of multiple BDTs.

Each BDT of the $\Kp$ ($\Kz$) ensemble uses~26 (38) discriminating variables, described in the Appendix.  
These variables fall into four general categories: quantities related to 
the missing energy in the event, to the overall event properties, to the signal kinematics, and to the overall reconstruction quality of the 
$\Btag$.  Some quantities are given in two different frames and thus allow the BDTs to extract from them additional discriminating power. Several additional variables were initially considered but were pruned during the BDT optimization process, as they 
were found to add little additional sensitivity.

``Missing Energy'' quantities relate to the fact that signal events are expected to possess significant missing energy 
and momentum because the signal decay includes two neutrinos.
In contrast, the dominant background events usually acquire missing energy and momentum as a result of particles passing outside of the detector fiducial acceptance, with the result that distributions of    quantities related to missing energy differ between signal and background.   

After the $\Btag$ and $K$ or $K_s^0$ signal candidate have been identified, signal events are expected to have little or no additional activity in the detector, other than a few low-energy clusters in the calorimeter  resulting from hadronic shower remnants, beam backgrounds, or similar sources. In contrast, background events arising from higher-multiplicity $B$ decays typically possess additional charged or neutral particles within the detector.  Variables which characterize this additional detector activity can provide discriminating power between signal and background, and are indicated by the term ``extra'' in the following.
  
The strongest discriminant for both $\Kp$ and $\Kz$ ensembles is $E_{\rm extra}$, the sum of all detector activity
not explicitly associated with either the $B_{\rm rec}$ or $K$ signal candidate, followed by 
$p^{*}_{\Kp}$ for the $\Kp$ ensemble and by the lab energy of the signal $\KS$ for the $\Kz$ ensemble. The
reconstructed mass of the $D$ from the $\Btag$ is the third ranking variable for both channels.

\begin{figure*}%[b!]
\begin{center}
\subfigure[]{\includegraphics[width=0.49\linewidth]{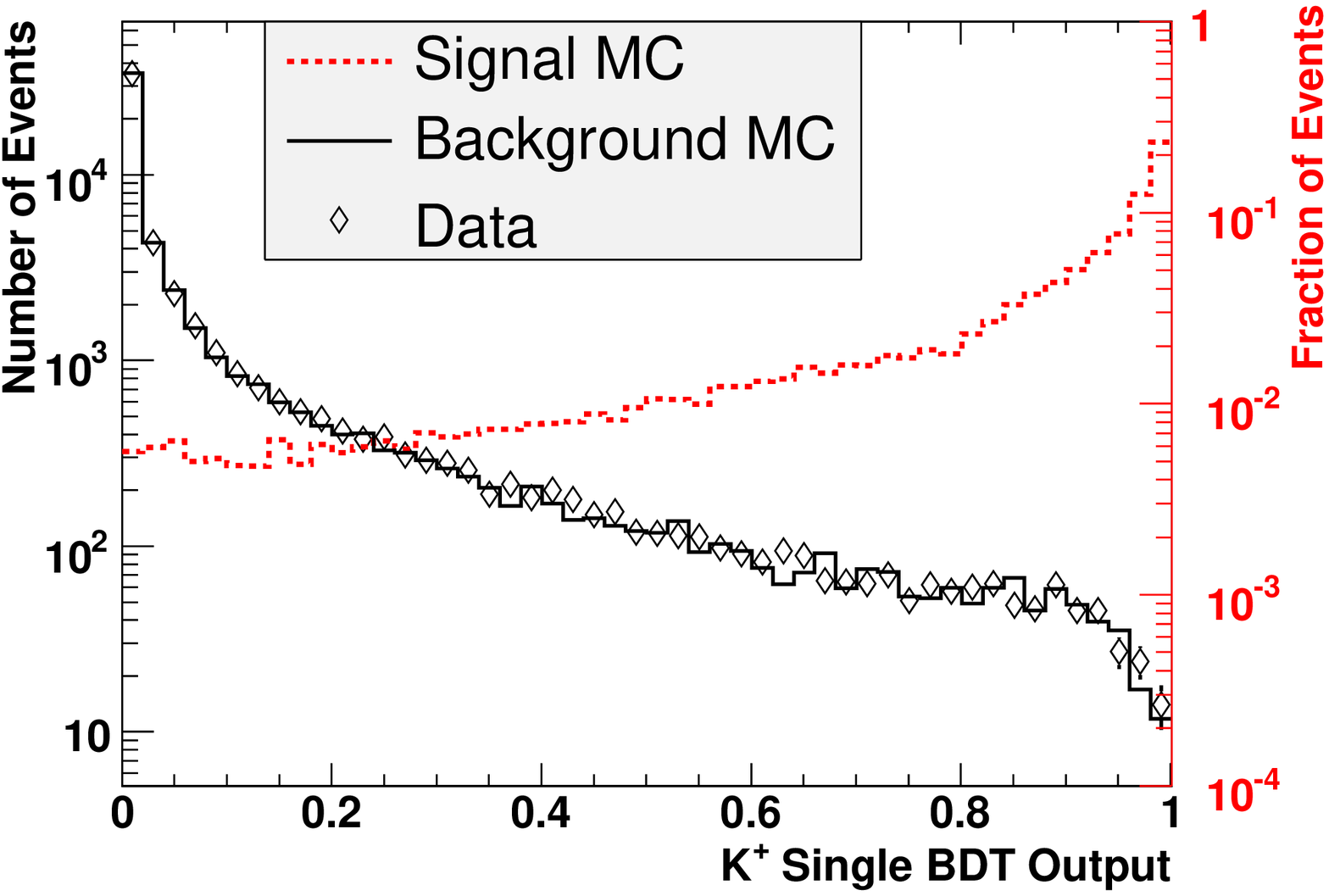}}\hfill
\subfigure[]{\includegraphics[width=0.49\linewidth]{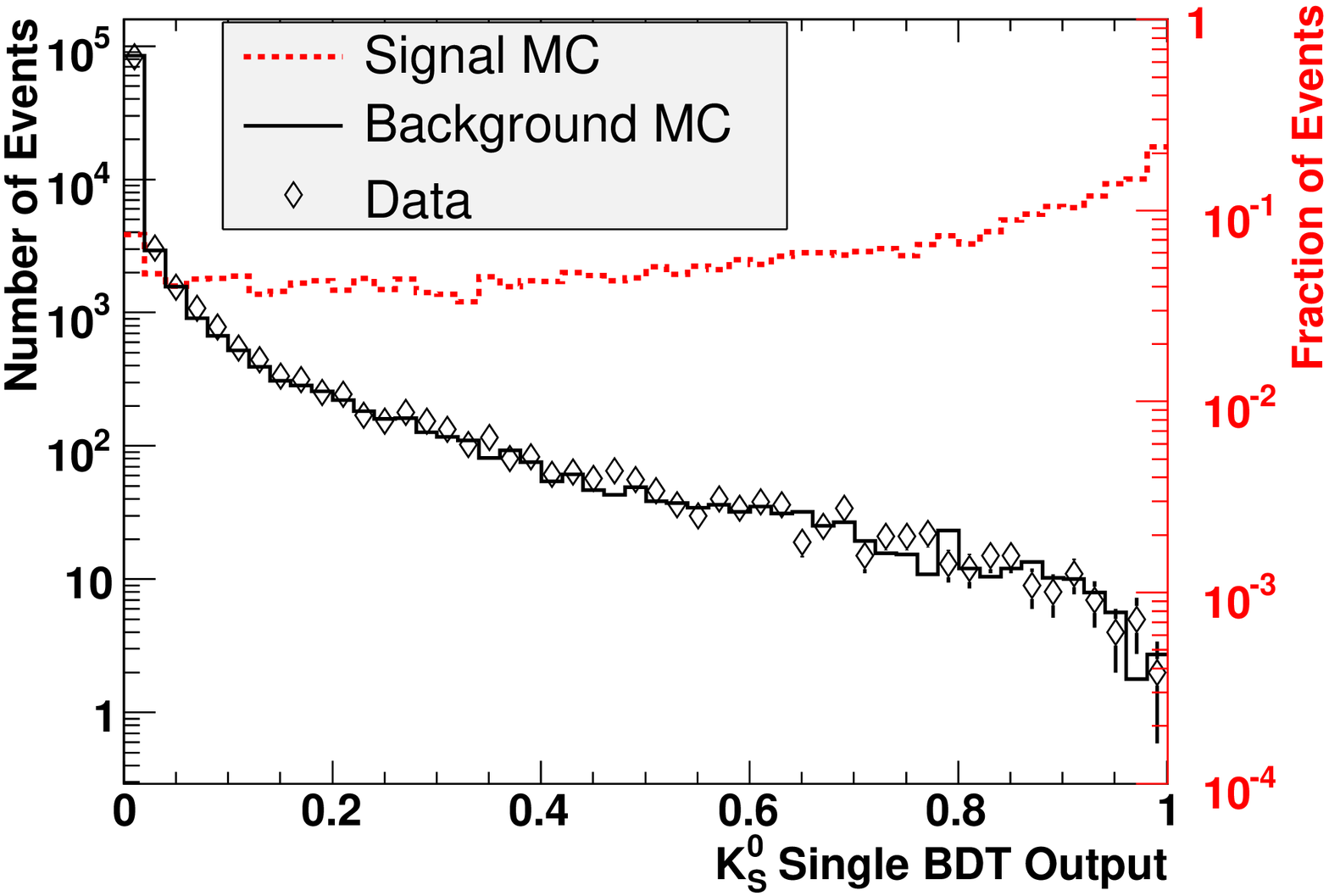}}
\caption{(a) $\Kp$ and (b) $\Kz$ BDT output for data (diamonds), background MC (solid), and  signal MC (dotted)   events.  For each plot, the scale for the data  and background events is on the left axis, and the scale for the signal events is on the right axis. The distribution of signal MC events  is normalized to unit area. }
\label{fig:bdtplotdat}
\end{center}
\end{figure*}

Figure~\ref{fig:bdtplotdat} shows signal, background, and data distributions from the validation set of $\Kp$ and $\Kz$ BDT output
for a BDT randomly selected from the 20 BDTs in the ensemble.
The other 19 BDTs are similar to that  shown.

We choose as the  target signal efficiency   the one that maximizes expected signal significance averaged over the 20 BDTs, under the assumption of a  branching fraction of  $3.8\times 10^{-6}$.
This signal significance is \(s/\sqrt{s+b}\), where $s$ is the number of signal events, and $b$ is the number of background events. Optimization using a figure of merit based upon signal efficiency and independent of assumed branching fraction yields similar results.  For each BDT, a BDT output value that
yields the target signal efficiency is calculated.  For example, the BDT output cuts for the BDTs shown in Figure \ref{fig:bdtplotdat} are 0.976 for the $\Kp$ BDT and 0.955 for the $\Kz$ BDT.  The mean background for target signal efficiency is obtained by averaging the individual background estimates from each of the 20 BDTs. Thus, we treat each ensemble of 20~BDTs as a set of correlated estimators for
the numbers of signal and
background events in a signal region defined by the target signal efficiency.

The low-$\mnn$ (high-$\mnn$) measurement uses the $\Kp$ ensemble but only includes events with $p^{*}_{\Kp}>1.5 \gevc$  ($p^{*}_{\Kp}<1.5\gevc$), which means that only those events are used to calculate the signal efficiency and the background prediction. The low-$\mnn$ measurement  has   the same BDT output cuts and background prediction as the primary $\Kp$ measurement, with only the signal efficiency changed by the restriction on  $p^{*}_{\Kp}$. On the other hand, the high-$\mnn$ measurement has its own set of BDT output cuts based upon its own optimized signal efficiency, along with its own background prediction.

\begin{figure*}
\begin{center}
\subfigure[]{\includegraphics[width=0.31\textwidth]{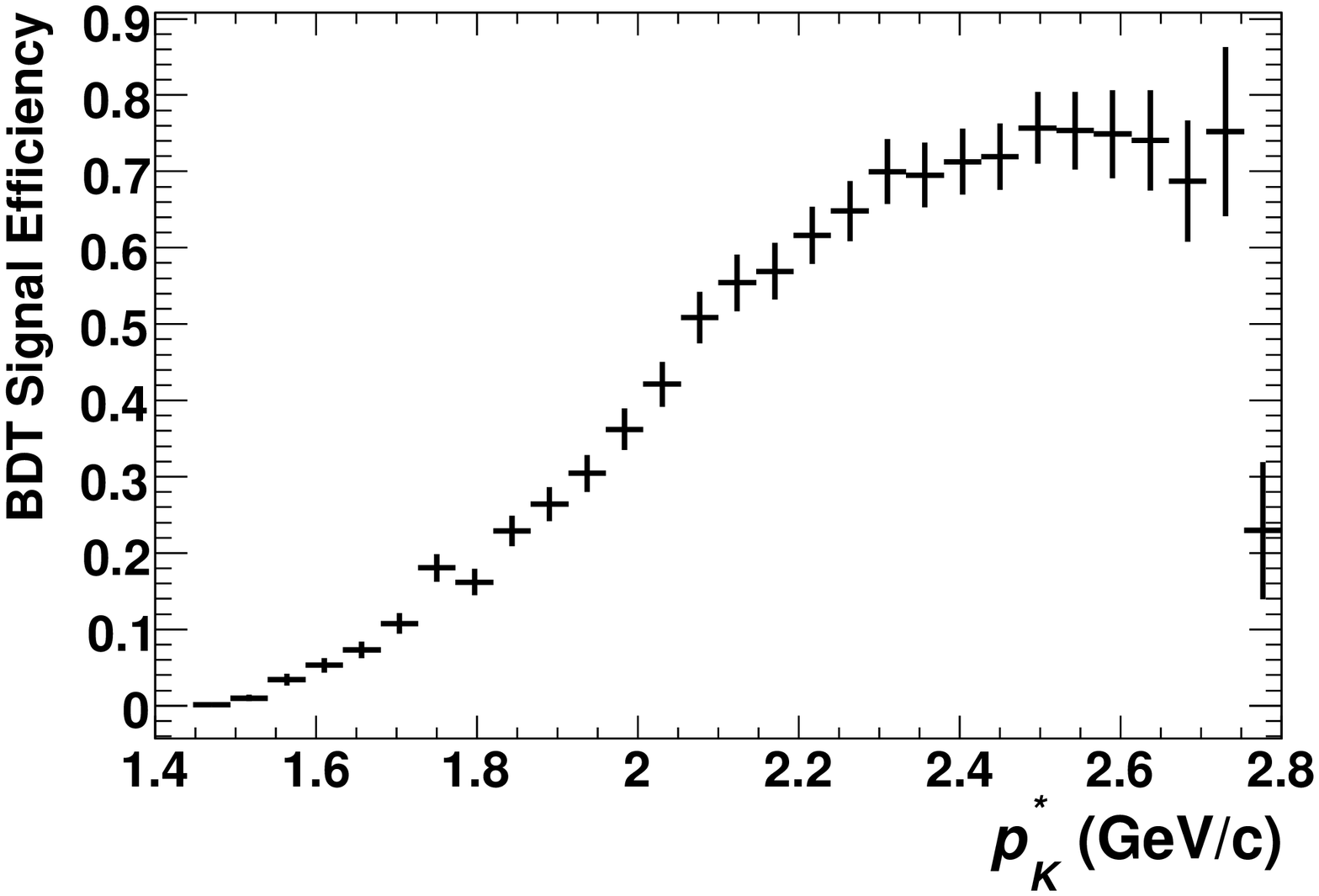}}
\subfigure[]{\includegraphics[width=0.31\textwidth]{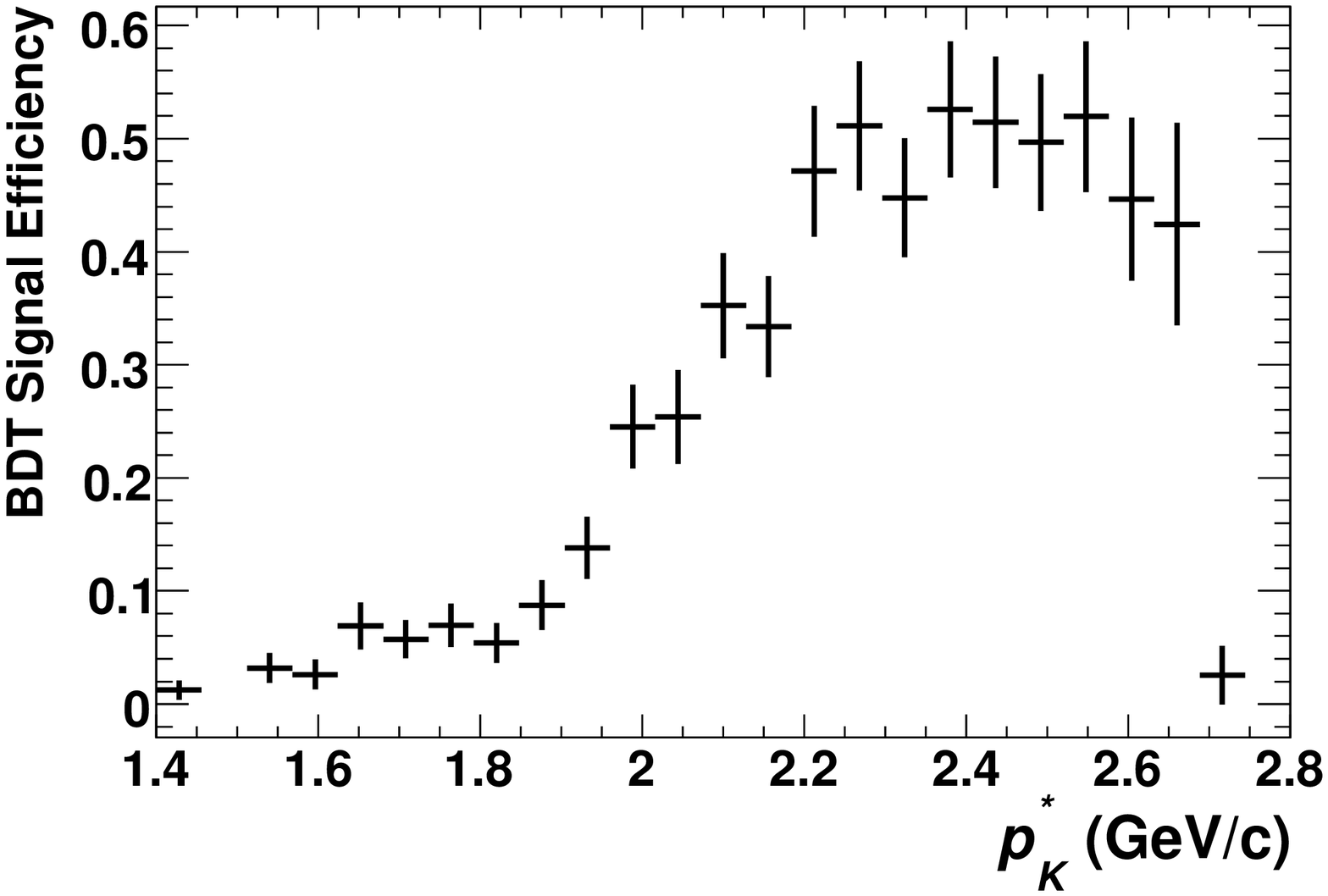}}
\subfigure[]{\includegraphics[width=0.31\textwidth]{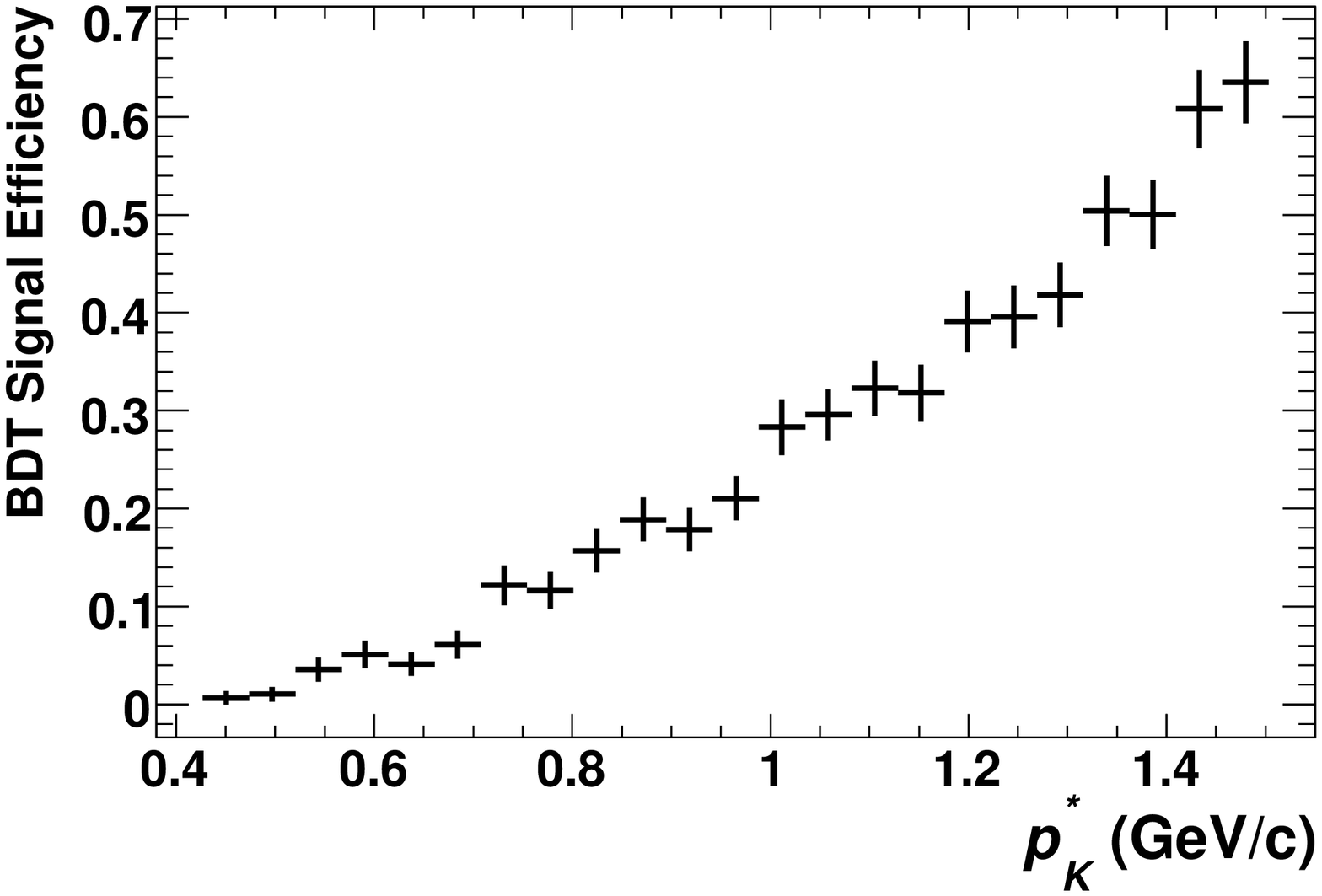}}
\caption{BDT selection efficiency in the signal region versus $p^{*}_{K}$ for (a) $\Kp$, (b) $\KS$ , and (c) high-$\mnn$ $\Kp$simulated signal events, considering only the effect of the BDT output cut.}
\label{fig:sigeff}
\end{center}
\end{figure*}

The total optimized signal 
efficiency for the $\Kp$ $(\Kz)$ mode is 0.16\% (0.06\%), while the
efficiency for the $\Kp$ low-$\mnn$ (high-$\mnn$) region is 0.24\% (0.28\%). The uncertainty in the signal efficiency is discussed below. Figure \ref{fig:sigeff} shows the BDT selection  efficiency versus $p^{*}_{K}$ for the $\Kp, \Kz,$ and high-$\mnn$ measurements, where the BDT selection efficiency considers only the effect of the BDT output cut. 

To measure the branching fractions, we  use the value obtained from simulated events of the predicted background in the signal region, the number of
observed data events, and the  signal efficiency, as shown by the following equation:  \(\mathcal{B}\ = (N_{obs}-N_{bkg})/\epsilon N_{B}\), where $\mathcal{B}$ is the branching fraction, $N_{obs}$ is the number of observed data events, $N_{bkg}$ is the number of predicted background events, $\epsilon$ is the total signal efficiency, and $N_{B}$ is the number of $B$ mesons, either charged or neutral~\cite{bbsame}, that are relevant  to the branching fraction. We account for the 50\% correlation between each of  the
datasets when computing the statistical uncertainty of the estimated
background contribution by using a standard method for combining correlated uncertainties~\cite{Barlow}.

Data control samples are used to ensure that both signal-like and
 background-like events in actual data are classified similarly to simulated
events. The vetoed $a_{1}^{+}$ events offer a high-statistics 
control sample which can be used to compare the $\Kp$ and $\Kz$ BDT output distributions
for background events in both simulated and actual data. We find good agreement
between data and MC events in the BDT output distribution for both final states, with only a $(+5 \pm 2)\%$ data-MC discrepancy.
For the \Kp mode we  make a $+5\%$ adjustment to the
expected number of background events, based upon  a weighting technique that corrects  data-MC discrepancy in the sideband $\Kp$ BDT output next to the signal region, and we assign the full adjustment as a systematic uncertainty.
Likewise, for the high-$\mnn$ \Kp measurement, we make a $+25\%$ correction to the
expected number of background events and assign the full correction as a systematic uncertainty. In the $\KS$ final state, we find a $(+10 \pm 3)\%$ data-MC discrepancy in the sideband
BDT output  next to the signal region, and we make a $+10\%$ correction
and assign the full correction as a systematic uncertainty. 

To validate our signal efficiency estimates
and assess their systematic uncertainties, we use high-purity samples of $\modekJpsill$ decays
(where $\ellell = \epem, \mumu$).  The
two leptons from the $\jpsi$ are discarded in order to model the unseen neutrinos of the signal decay, and then the events are subjected to the same selection requirements as other signal candidates. 
Classifying $\jpsi K$ data and MC events, we find only a $(-10 \pm 10)\%$ data-MC discrepancy  in the BDT output distribution. Although we do not make any correction, we assign
a $10\%$ systematic uncertainty to the estimated signal efficiency for all four measurements ($\Kp, \KS,$ \mbox{low-$\mnn \Kp,$} high-$\mnn  \Kp$)  based
on these results. We also assign a signal efficiency systematic uncertainty of $10\%$
to account for the theoretical uncertainties of the  signal  models. Adding these in quadrature, we assign a total
uncertainty of $14\%$ in the estimation of signal efficiency for both
final states.
Table \ref{tab:systematics} summarizes all of the systematic uncertainties.

\begin{table}
%\footnotesize 
\centering
\caption{Systematic uncertainties}
\begin{tabular}{lc}
\hline \hline
Category            & Uncertainty    \\ \hline %\\
Signal efficiency    & $14\%$  \\
$\Kp$ background prediction     & $5\%$   \\
High-$\mnn$ $\Kp$ background prediction    & $25\%$   \\
$\KS$ background prediction     & $10\%$   \\
\hline \hline
\end{tabular}
\label{tab:systematics}
\end{table}

Table~\ref{tab:nexp} shows the total signal efficiencies and the expected number of   signal and
background events in the data. We performed a
blind analysis where data events with BDT outputs above the optimized values
were not counted or plotted until the analysis methodology and sources of systematic
uncertainty were fixed as described above.

\begin{table}
%\footnotesize 
\centering
\caption{Total signal efficiencies and MC expectations of the number of data events.
The uncertainties  shown are systematic for $N_{\rm{sgnl}}$, with statistical  negligible, and statistical followed by systematic  for
$N_{\rm{bkgd}}$.}
\begin{tabular}{lccc}
\hline \hline
Mode & $\epsilon$ (in \%)  & $N_{\rm{sgnl}}$ & $N_{\rm{bkgd}}$ \\ \hline %\\
$\Kp$          &  0.16 & $2.9 \pm 0.4 $ & $17.6 \pm 2.6 \pm 0.9$ \\
$\KS$          & 0.06 & $0.5 \pm 0.1 $ & $3.9  \pm 1.3 \pm 0.4$ \\
low-$\mnn$ $\Kp$ & 0.24 & $2.9 \pm 0.4 $ & $17.6 \pm 2.6 \pm 0.9$ \\
high-$\mnn$ $\Kp$ & 0.28 & $2.1 \pm 0.3 $ & $187  \pm 10  \pm 46$  \\
\hline \hline
\end{tabular}
\label{tab:nexp}
\end{table}

\begin{table}
\centering
\caption{ Observed and excess data events, with statistical uncertainties \cite{Barlow2} shown for $N_{\rm{obs}}$ and combined statistical and systematic uncertainties shown for $N_{\rm{excess}}$. The last column shows
the probability that excess events could be due solely  to a
background fluctuation.}
\begin{tabular}{lccc}
\hline \hline
Mode          & $N_{\rm{obs}}$      & $N_{\rm{excess}}$   & Prob. \\ \hline %\\
$\Kp$         & $19.4^{+4.4}_{-4.4}$      & $1.8^{+6.2}_{-5.1}$ & 38\% \\
$\Kz$         & $6.1^{+4.0}_{-2.2}$ & $2.2^{+4.1}_{-2.8}$ & 23\% \\
low-$\mnn$ $\Kp$   & $19.4^{+4.4}_{-4.4}$      & $1.8^{+6.2}_{-5.1}$ & 38\% \\
high-$\mnn$ $\Kp$  & $164^{+13}_{-13}$        & $-23^{+49}_{-48}$   & 33\% \\
\hline \hline
\end{tabular}
\label{tab:obs}
\end{table}

\begin{table}
\centering
\caption{Branching fraction (BF) central values and upper limits. The low- and high-$\mnn$ values are partial BFs, while the rest are total BFs.}
\begin{tabular}{lccc}
\hline \hline
Mode            & BF    & 90\% CL  & 95\% CL   \\ 
    &  $\times 10^{-5}$    &  $\times 10^{-5}$  &  $\times 10^{-5}$  \\ 
\hline
$\Kp$           & $0.2^{+0.8}_{-0.7}$  & 1.3   & 1.6  \\
$\Kz$           & $1.7^{+3.1}_{-2.1}$     & 5.6   & 6.7  \\
Comb. $\Kp,\Kz$ &  $0.5^{+0.7}_{-0.7}$  & 1.4   & 1.7  \\
Low-$\mnn$ $\Kp$    & $0.2^{+0.6}_{-0.5}$  & 0.9   & 1.1  \\
High-$\mnn$ $\Kp$     & $-1.8^{+3.8}_{-3.8}$    & 3.1  & 4.6  \\
\hline \hline
\end{tabular}
\label{tab:results}
\end{table}

\begin{figure*}
\begin{center}
\subfigure[]{\includegraphics[width=0.32\textwidth, trim= 0.0in 0 0.1in 0]{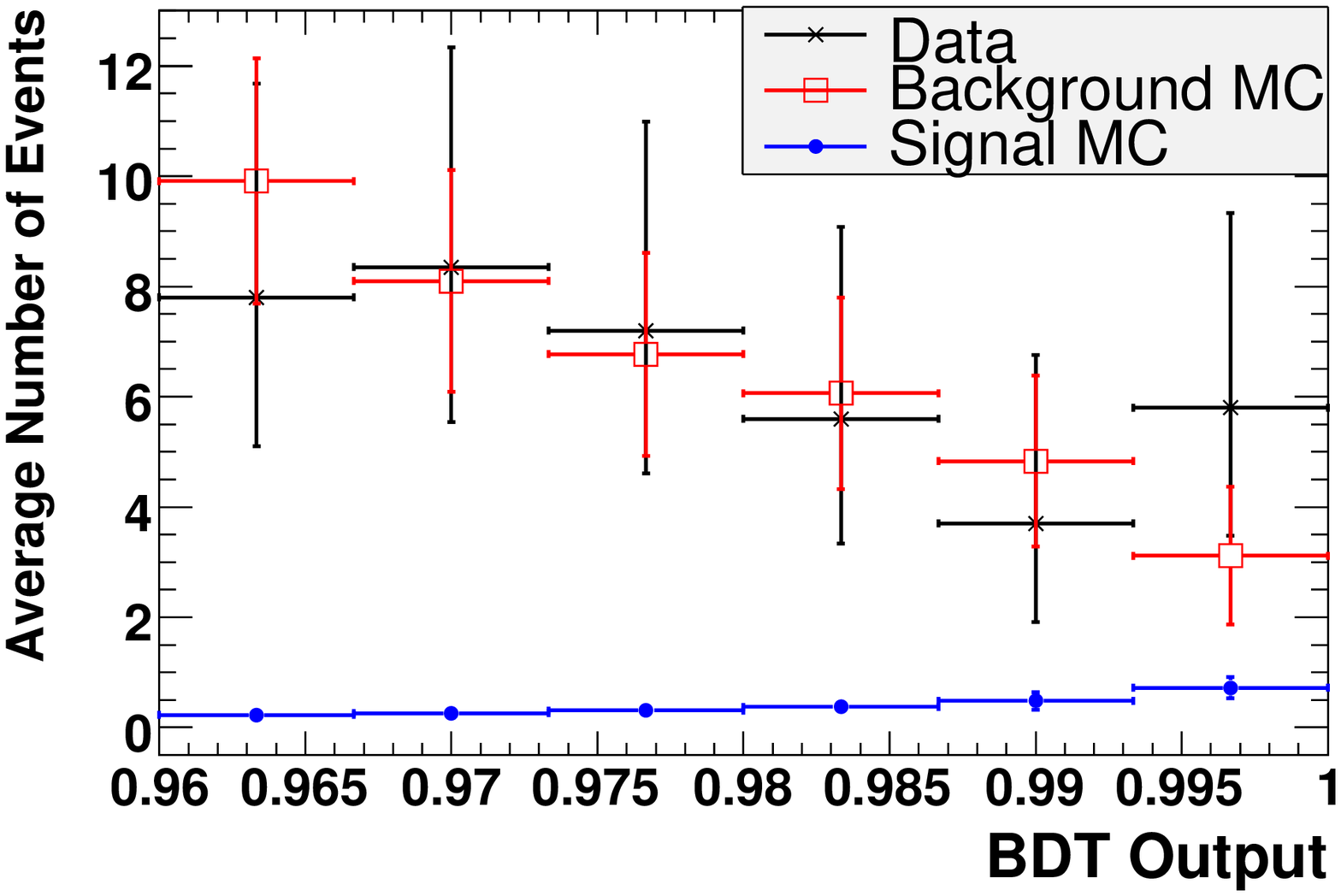}}
\subfigure[]{\includegraphics[width=0.32\textwidth, trim= 0.0in 0 0.1in 0]{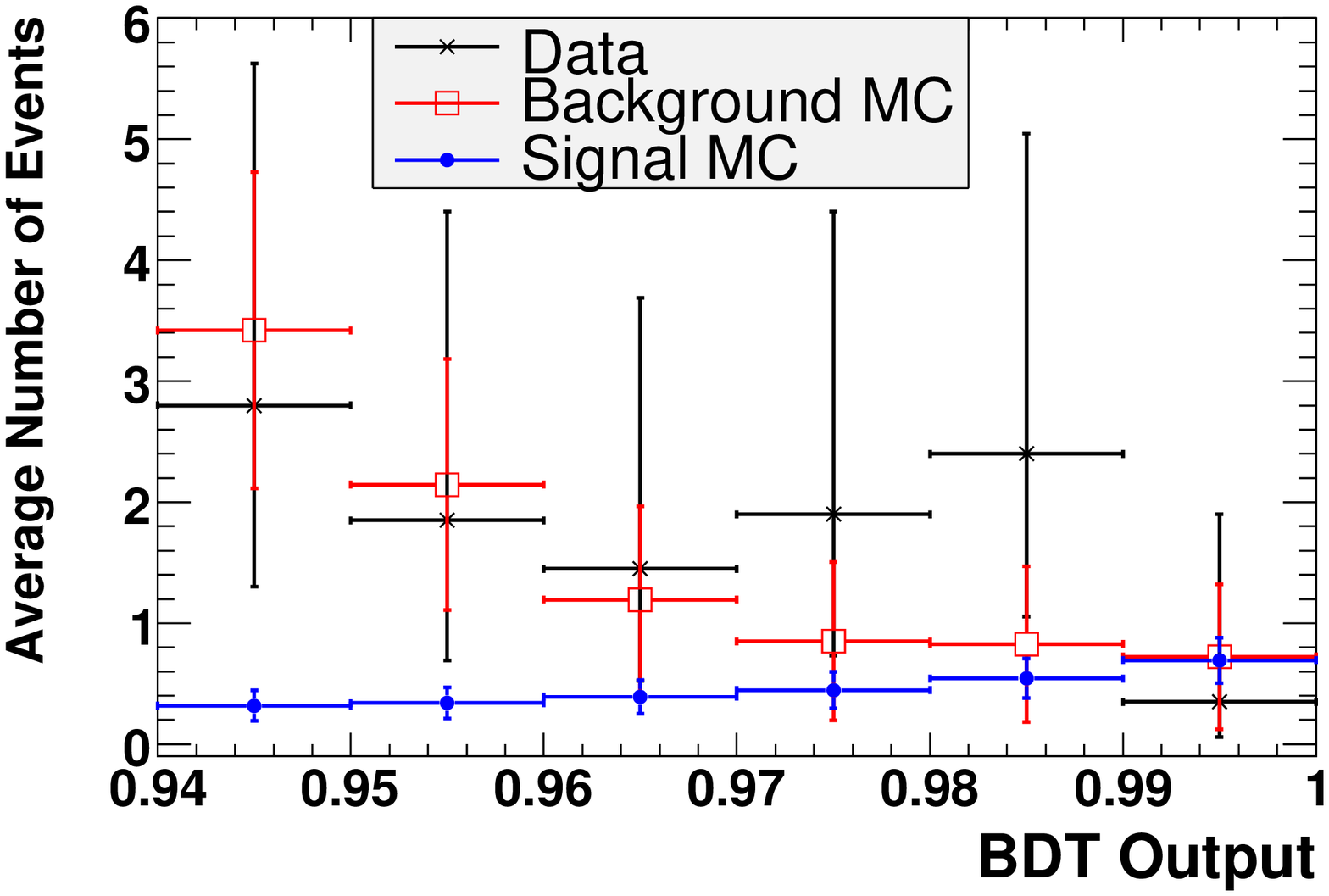}}
\subfigure[]{\includegraphics[width=0.32\textwidth, trim= 0.0in 0 0.1in 0]{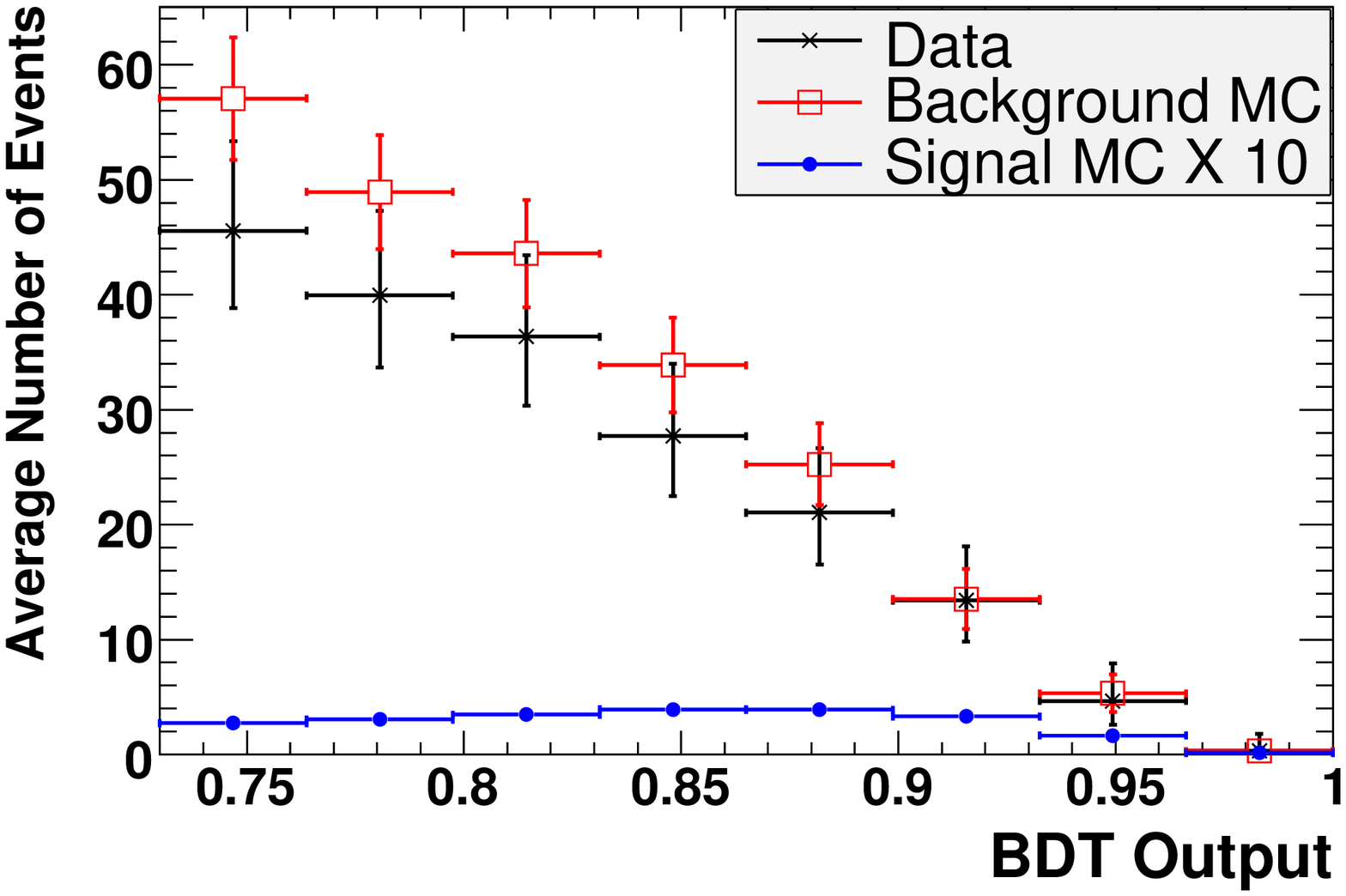}}
\caption{Averaged BDT signal-region output for (a)  $\Kp$, (b) $\KS$, and (c) high-$\mnn$ $\Kp$ data, with expected signal and background contributions. The signal estimate assumes a  branching fraction of  $3.8\times 10^{-6}$. }
\label{fig:datamcsigreg}
\end{center}
\end{figure*}

\begin{figure*}
\begin{center}
\subfigure[]{\includegraphics[width=0.32\textwidth, trim= 0in 0 0.2in 0]{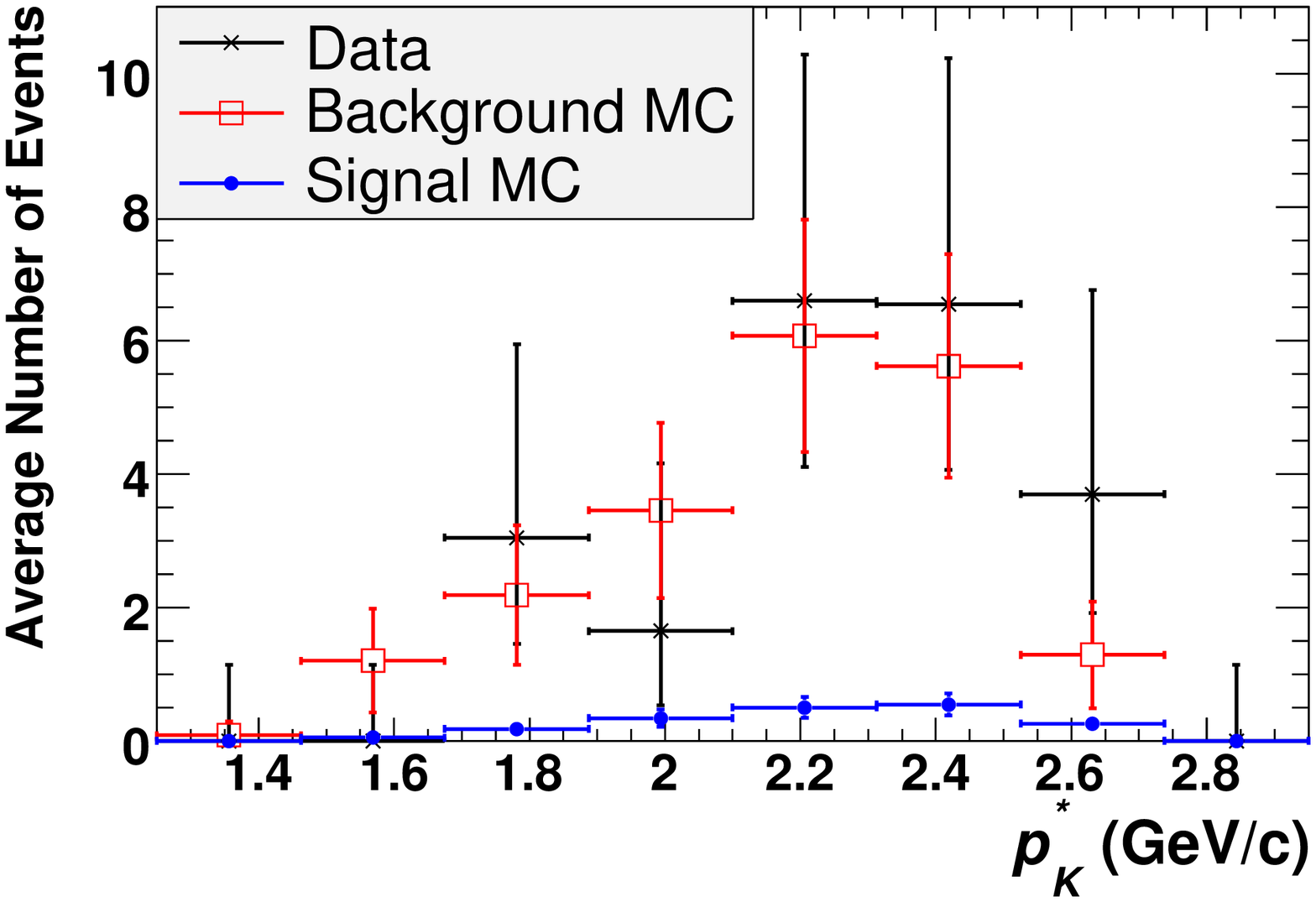}}
\subfigure[]{\includegraphics[width=0.32\textwidth, trim= 0in 0 0.2in 0]{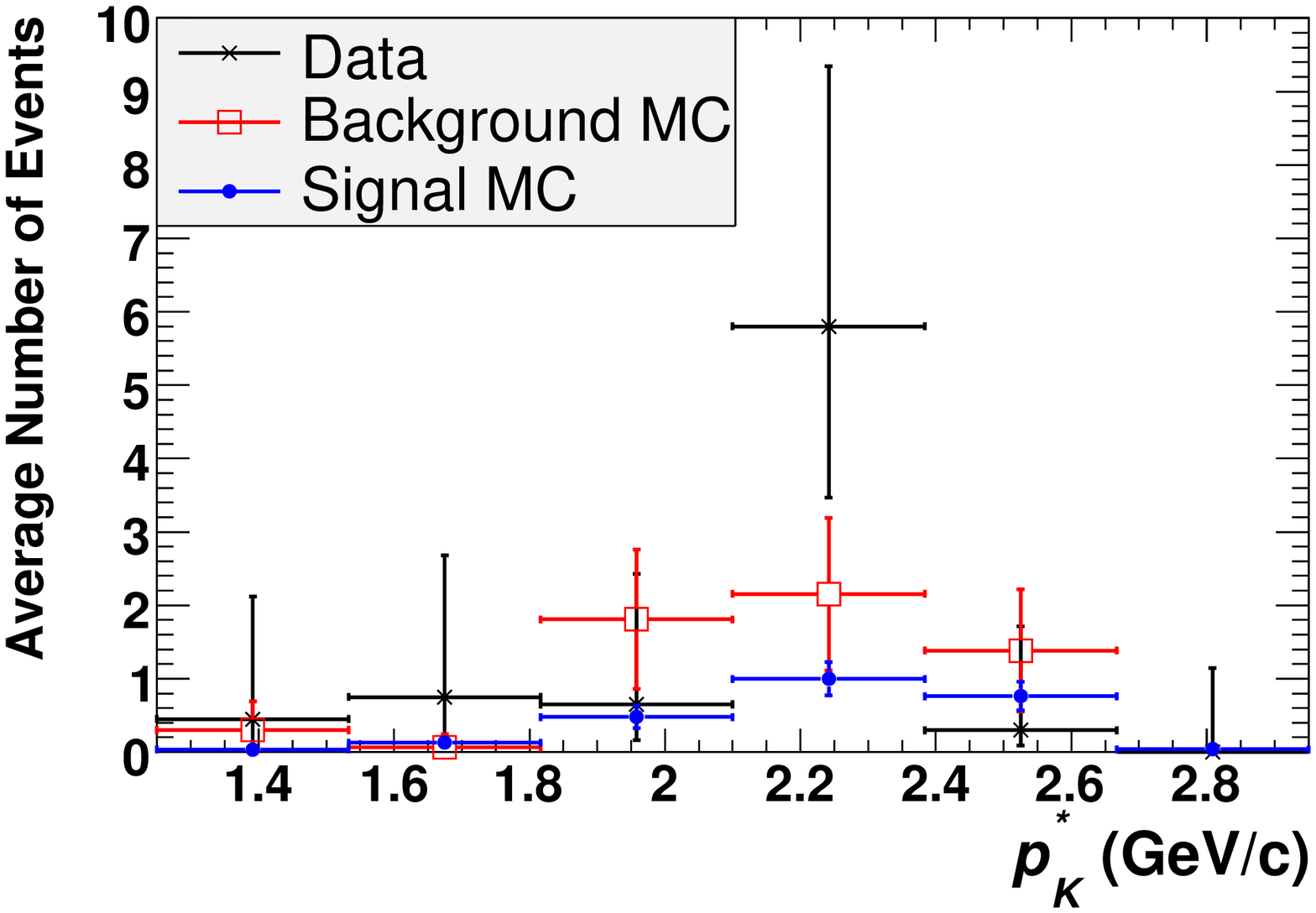}}
\subfigure[]{\includegraphics[width=0.32\textwidth, trim= 0in 0 0.2in 0]{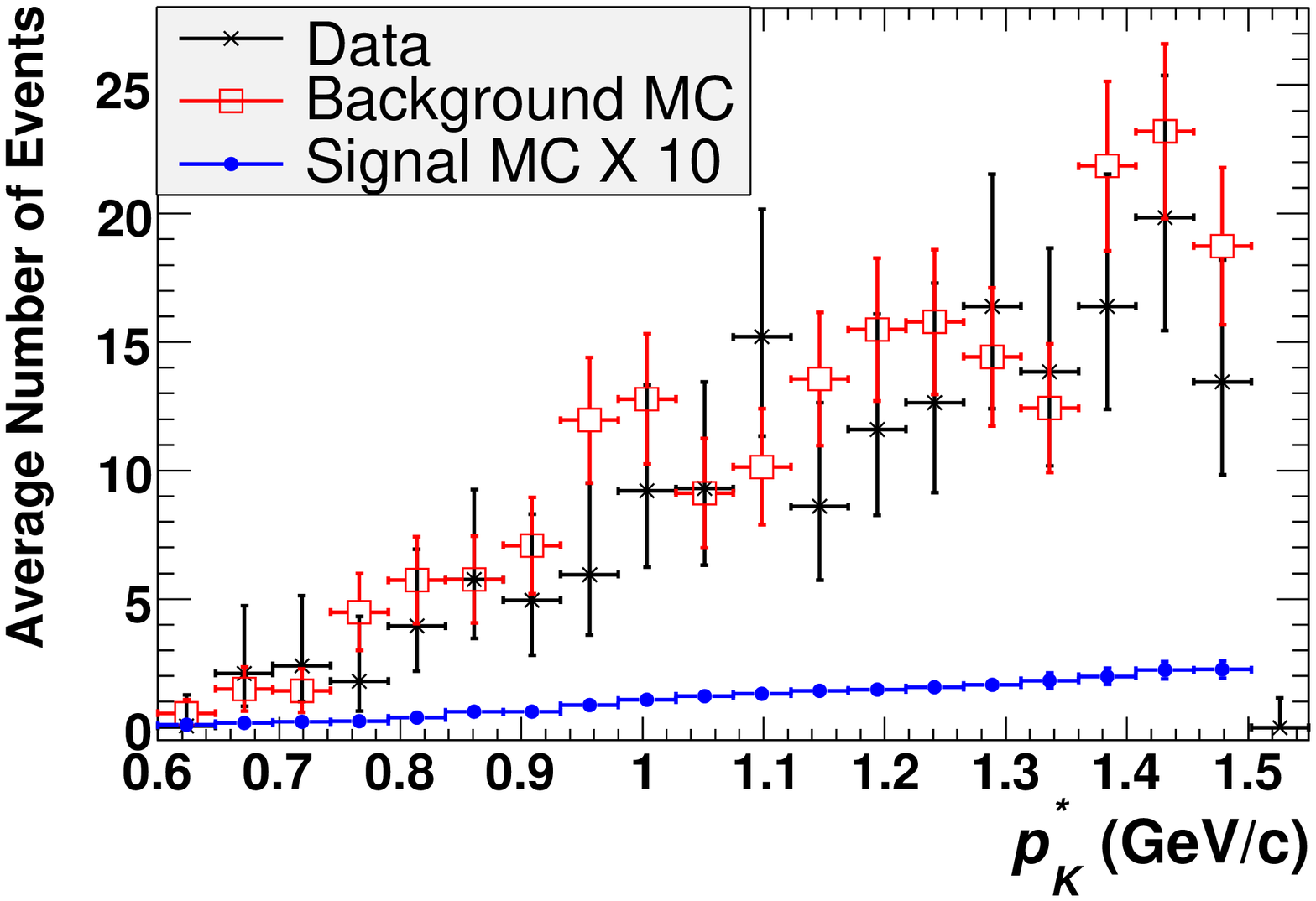}}
\caption{Averaged $p^{*}_{K}$ signal-region output for (a)  $\Kp$, (b) $\KS$ , and   (c) high-$\mnn$ $\Kp$ data,
with expected signal and background contributions. The signal estimate assumes a  branching fraction of  $3.8\times 10^{-6}$. }
\label{fig:kp3cmsigreg}
\end{center}
\end{figure*}

\begin{figure*}[htb!]
\begin{center}
\subfigure[]{\includegraphics[width=0.32\textwidth]{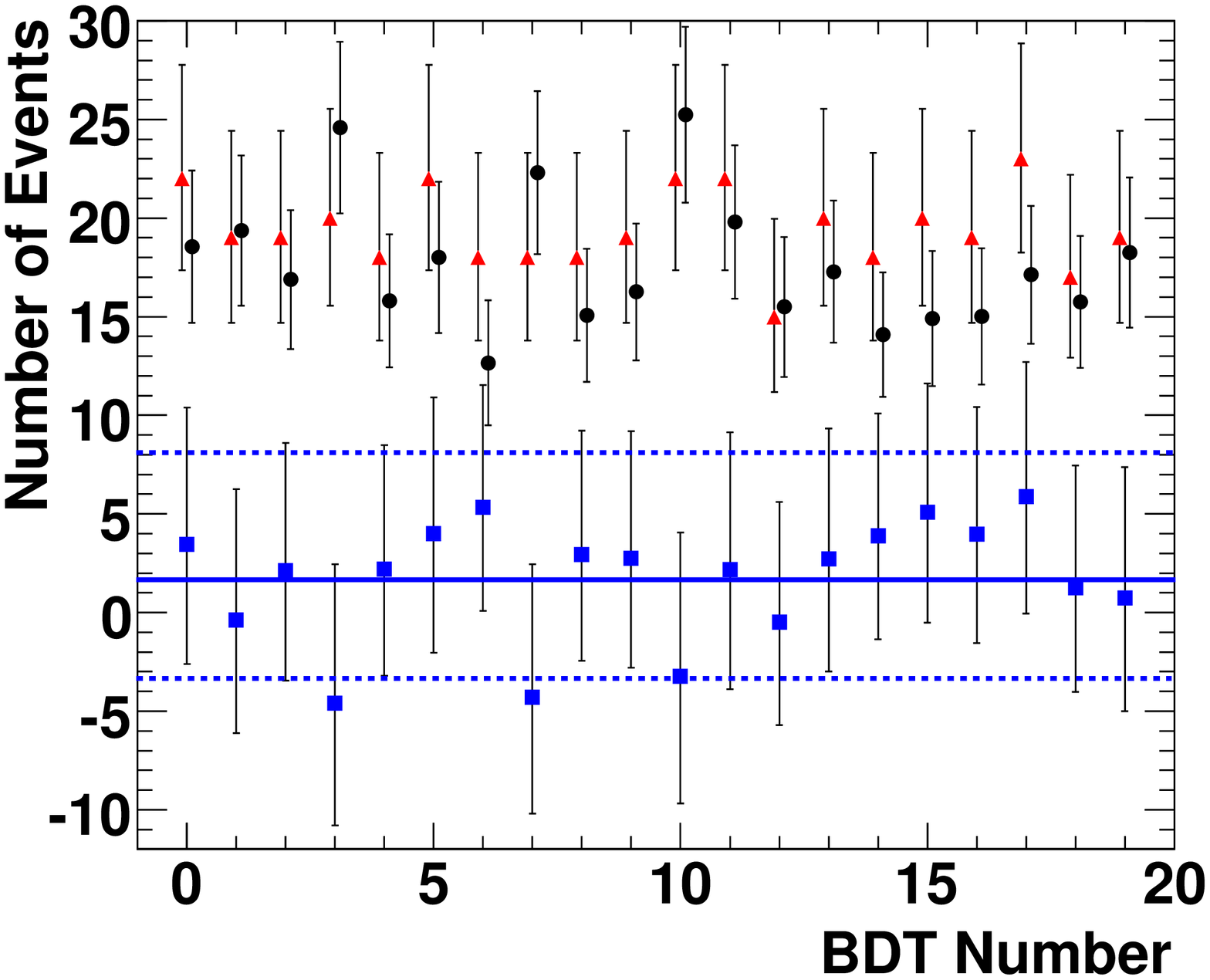}}
\subfigure[]{\includegraphics[width=0.32\textwidth]{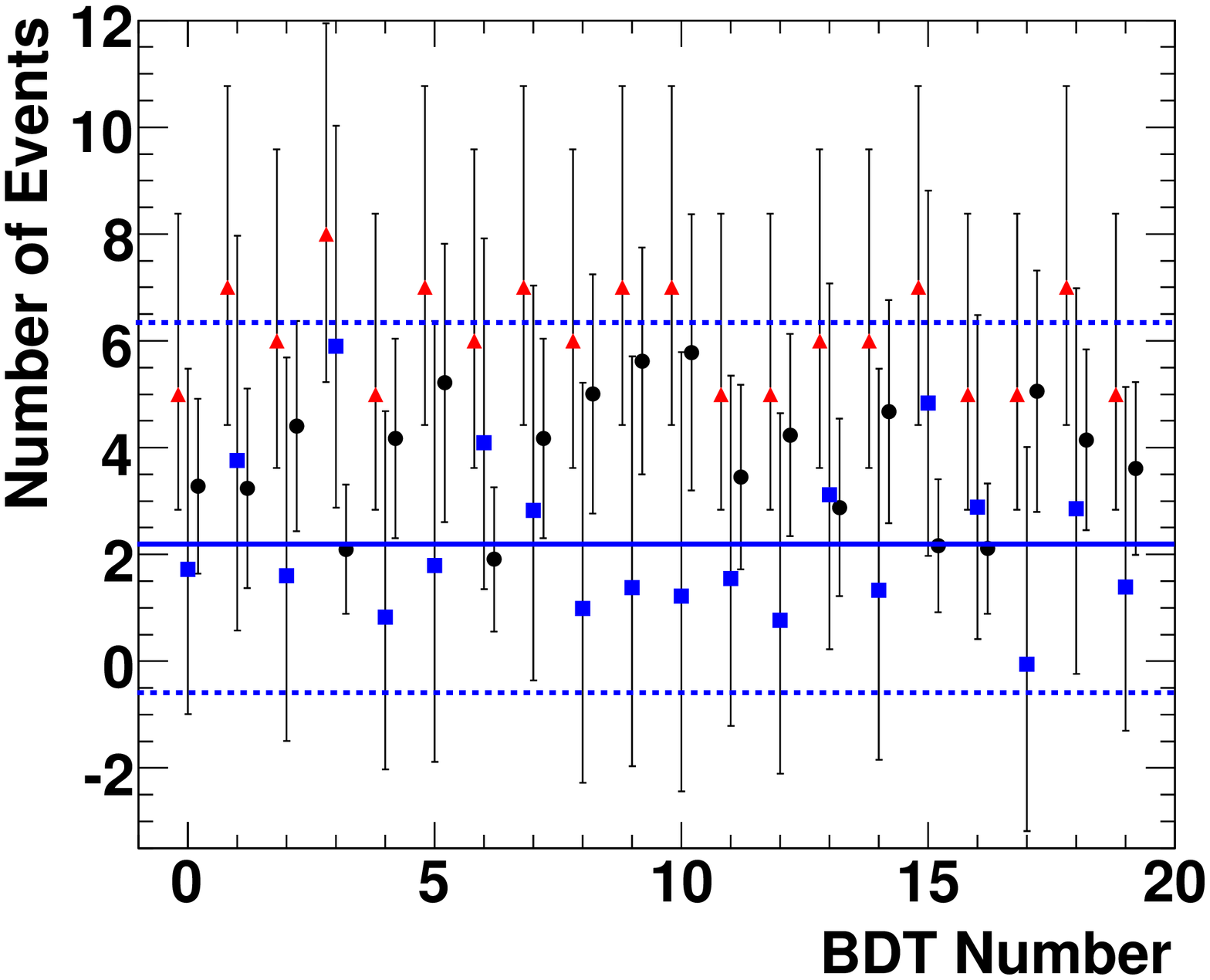}}
\subfigure[]{\includegraphics[width=0.32\textwidth]{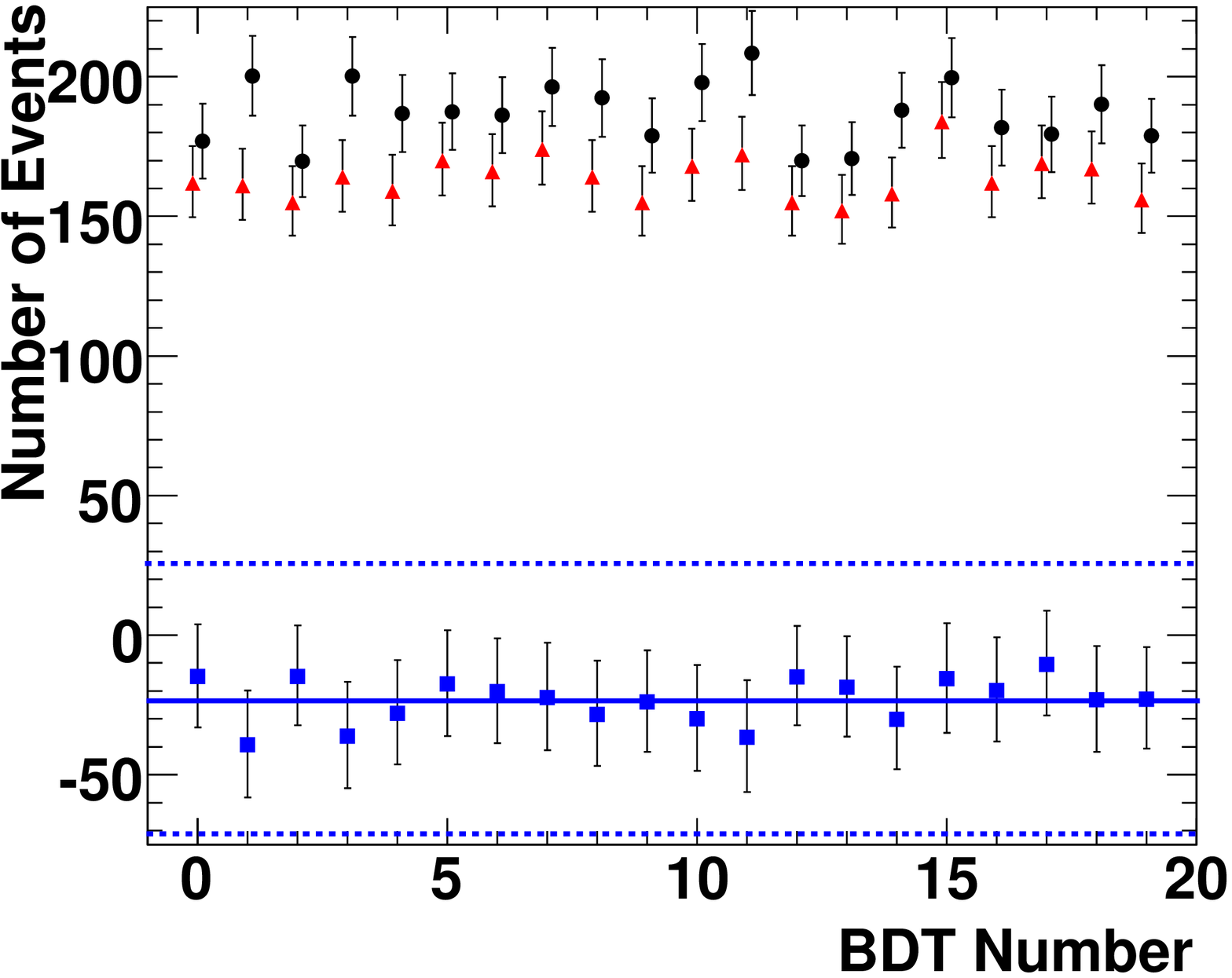}}
\caption{Integrated numbers of observed (red triangles), expected background (black circles), and
excess events (blue squares)  for data for each  BDT:
(a)  $\Kp$, (b)  $\KS$, and (c)  high-$\mnn$ $\Kp$. The individual uncertainties are purely statistical and assume no correlation between data sets.
The horizontal dashed lines show the sum of the statistical and systematic  uncertainties
on the mean number of excess events.}
\label{fig:bytreesigreg}
\end{center}
\end{figure*}

Table~\ref{tab:obs} shows our results. The non-integer number of
observed events results from averaging the integer yields from
the 20  BDTs of each type. We calculate two-sided $68\%$
confidence intervals for the number of excess events
based on the statistical and systematic
uncertainties in the background estimates, as well as the
statistical uncertainty on the number of events observed in
the data. Figure~\ref{fig:datamcsigreg}
shows the averaged BDT outputs in the signal region for $\Kp$, $\Kz$, and high-$\mnn$ $\Kp$ data
overlaid with the background and signal contributions, while Figure~\ref{fig:kp3cmsigreg} shows similar plots for the $p^{*}_{K}$ distribution in the signal region. Figure~\ref{fig:bytreesigreg} shows the integrated numbers of  events (observed, predicted background, and excess over background) in the signal region for $\Kp$, $\Kz$, and high-$\mnn$ $\Kp$ data
for each of the 20 BDTs of each type.
Table~\ref{tab:results} gives the branching fraction central values, along with corresponding 90\% and 95\% CL upper limits,
assuming the BHI signal model (the ABSW model gives similar results). The upper limits are calculated using a frequentist method~\cite{Barlow2}. The quoted uncertainties include all statistical and systematic uncertainties.
Our results  constrain the
$\modekavgnn$ branching fraction at the 90\% CL to a few times the SM expectation, with limits of $1.3 \times 10^{-5}$ for \modekplnn
and $5.6 \times 10^{-5}$ for $\modekzeroallnn$. 

We are grateful for the 
extraordinary contributions of our \pep2\ colleagues in
achieving the excellent luminosity and machine conditions
that have made this work possible.
The success of this project also relies critically on the 
expertise and dedication of the computing organizations that 
support \babar.
The collaborating institutions wish to thank 
SLAC for its support and the kind hospitality extended to them. 
This work is supported by the
US Department of Energy
and National Science Foundation, the
Natural Sciences and Engineering Research Council (Canada),
the Commissariat \`a l'Energie Atomique and
Institut National de Physique Nucl\'eaire et de Physique des Particules
(France), the
Bundesministerium f\"ur Bildung und Forschung and
Deutsche Forschungsgemeinschaft
(Germany), the
Istituto Nazionale di Fisica Nucleare (Italy),
the Foundation for Fundamental Research on Matter (The Netherlands),
the Research Council of Norway, the
Ministry of Education and Science of the Russian Federation, 
Ministerio de Ciencia e Innovaci\'on (Spain), and the
Science and Technology Facilities Council (United Kingdom).
Individuals have received support from 
the Marie-Curie IEF program (European Union), the A. P. Sloan Foundation (USA) 
and the Binational Science Foundation (USA-Israel).

%\clearpage
\textbf{\begin{center} Appendix \end{center}}

\noindent \begin{center} \textbf{Definitions of BDT Variables} \end{center} 
In the following the notation $[\Kp]$ or  $[\Kz]$ indicates that a variable is used only by that ensemble; otherwise it is used by both BDT ensembles.
\\ 

\noindent \uline{BDT input variables related to missing \mbox{4-momentum}} \\    The event missing 4-momentum is 
computed from the difference between the 4-momentum of the combined $e^{+}e^{-}$ beams and the 4-momenta of all 
 charged and neutral particles reconstructed in the detector. 

\begin{itemize}
\item Energy component of missing momentum 4-vector
 
\item Energy component of missing momentum 4-vector (CMS)
  
\item Magnitude of the missing momentum 3-vector 

\item Magnitude of the missing momentum 3-vector (CMS) 

\item Cosine of the angle with respect to the beam axis
of the missing momentum 3-vector 

\item Cosine of the angle with respect to the beam axis 
of the 3-momentum vector representing the
difference between the initial event 
momentum and the summed momenta of the $\Btag$ and $\Bsig$ 
candidates  $[\Kz]$  

\end{itemize}

\noindent \uline{BDT input variables related to overall event properties} 

\begin{itemize}

\item $\Eex = \Sigma_i E_i$, where $E_i$ is the energy of an isolated 
EMC cluster or a charged track and the sum is over all 
tracks or clusters which are not part of the $\Btag$
or the $\Bsig$ 

\item Total energy of all reconstructed charged and neutral
particles in the event  

\item Minimum invariant mass obtained from the combination 
 of any three charged tracks in the event

\item Total charge of all tracks in the event $[\Kz]$

\item Total charge of all tracks matched to EMC energy deposits $[\Kz]$

\item Number of extra EMC clusters 

\item Number of $\K_L$  candidates in the EMC
 
\item Number of IFR  $\K_L$ candidates $[\Kp]$ 
 
\item Number of extra reconstructed tracks 

\item Magnitude of the 3-momentum of a candidate  $\Upsilon(4S)$ 
computed from the $\Btag$ and $\Bsig$ \mbox{4-momenta} $[\Kz]$ 

\item Angle with respect  to the beam axis of a candidate 
 $\Upsilon(4S)$ 3-momentum vector  computed from the $\Btag$ and $\Bsig$ 4-momenta $[\Kz]$ 

\item Normalized second Fox-Wolfram moment of the overall event  

\end{itemize}

\noindent \uline{BDT input variables related to signal kinematics}  
\begin{itemize}

\item Cosine of the angle between the signal $K$  
and the event thrust axis

\item Cosine of the angle between the signal $K$ 
and the $Dl$ thrust axis

\item Energy of the signal kaon $[\Kz]$ 

\item Reconstructed invariant mass of the signal $\KS$  $[\Kz]$ 

\item Magnitude of the 3-momentum of the signal kaon 

\item Magnitude of the CMS 3-momentum of the signal kaon 

\item Cosine of the angle with respect to the beam axis of the 3-momentum vector of the signal kaon 

\item Uncertainty in the $x$-component of the signal $K$ point of  
 closest approach to the $e^{+}e^{-}$ interaction point, as determined from a three dimensional fit, with the $x$-axis defined  perpendicular to the beam axis in the horizontal plane of the detector $[\Kz]$ 

\item Uncertainty in the $x$-component of the signal $K$ point of closest approach to the $e^{+}e^{-}$ interaction point, as determined by a fit in the $xy$-plane, with the  $x$-axis defined  perpendicular to the beam axis~ ($z$)  in the horizontal plane of the detector  $[\Kz]$

\end{itemize}

\noindent \uline{BDT input variables related to $\Btag$ reconstruction}  

\begin{itemize}

\item $\chi^2$ per degree of freedom of the vertex fit of the tracks making up the $\Btag$ 

\item $ \cos\theta_{BY} \equiv $   
$ (2E^{*}_{beam}\cdot E^{*}_{Dl}-m^2_{Bn}-m^2_{Dl})/ (2 p^{*}_{Dl}\ \cdot\sqrt{E^{*2}_{beam}-m^2_{Bn}}) $, where $E^{*}_{beam}$ is one half the total CMS energy, $m_{Bn}$ is the nominal $B$ meson mass~\cite{pdg:2008} and $E^{*}_{Dl}$, $m_{Dl}$ and $p^{*}_{Dl}$ are the CMS energy, invariant mass and 3-momentum magnitude of the $D$ - lepton combination used in the reconstruction of the $\Btag$ 

\item $\cos\theta_{BY}$ re-calculated with the addition of a photon to 
the $Dl\nu$ candidate such that $100 < (m(\Dz,\gamma) - m(\Dz)) < 150\mevcc$ 

\item Reconstructed decay mode of the $D$ from the $\Btag$ 

\item Uncertainty in the $x$-component of the 
point of closest approach to the $e^{+}e^{-}$ interaction point of the leading pion daughter from the $D$ meson, with the $x$-axis defined  perpendicular to the beam axis in the horizontal plane of the detector

\item Number of daughters possessed by the reconstructed $D$ from the $\Btag$

\item Number of extra $\pi^{0}$ candidates satisfying $ 0.115 < m(\gamma \gamma) <  0.150 $\gevcc and $E_\gamma >30$ MeV

\item Reconstructed invariant mass of the $\Btag$ 

\item Reconstructed invariant mass of the $D$ from the $\Btag$

\item Magnitude of the CMS 3-momentum of the $\Btag$  $[\Kz]$

\item Magnitude of the CMS 3-momentum of the $D$ candidate from the $\Btag$

\item Magnitude of the 3-momentum of the lepton from the $\Btag$ $[\Kz]$

\item Magnitude of the CMS 3-momentum of the lepton from the $\Btag$

\end{itemize}

\textit{}


\begin{thebibliography}{99}
\bibitem{asterisk}
All kinematic quantities in this paper are defined in the lab frame unless 
marked with an asterisk, in which case they are in the CMS.
\bibitem{bbsame}
Equal branching fractions for $\upsbzbz$  and $\upsbpbm$  are assumed in this paper.
\end{thebibliography}
\end{document}